\documentclass[
reprint,
superscriptaddress,
amsmath,
amssymb,
aps,
prb,
]{revtex4-2}
\usepackage{ragged2e}
\usepackage{caption}
\DeclareCaptionJustification{justified}{\justifying}
\usepackage[colorlinks,allcolors=cyan]{hyperref}
 \newcommand{\av}[1]{\left\langle #1 \right \rangle}
\usepackage{graphicx}
\usepackage{dcolumn}
\usepackage{bm}
\usepackage[capitalise]{cleveref}
\usepackage{physics}
\usepackage{xcolor}
\usepackage{comment}
\usepackage{bbold}
\usepackage{mathtools}
\usepackage{bbold}
\usepackage[caption=false]{subfig}
\DeclareCaptionOption{justified}[]{\caption@setjustification{justified}}

\begin{document}
\def\lc{\left\lfloor}   
\def\rc{\right\rfloor}
\setlength{\intextsep}{10pt plus 2pt minus 2pt}

\title{Complex path simulations of geometrically frustrated ladders}
 
\author{Elyasaf Y. Cohen}
 \affiliation{Racah Institute of Physics, The Hebrew University of Jerusalem, Jerusalem 91904, Israel}
 \author{Andrei Alexandru}
 \affiliation{Department of Physics, The George Washington University,
Washington, DC 20052, U.S.A.}
 \author{Snir Gazit}
 \affiliation{Racah Institute of Physics, The Hebrew University of Jerusalem, Jerusalem 91904, Israel}
 \affiliation{The Fritz Haber Research Center for Molecular Dynamics, The Hebrew University of Jerusalem, Jerusalem 91904, Israel}
\date{\today}

\begin{abstract}
Quantum systems with geometrical frustration remain an outstanding challenge for numerical simulations due to the infamous numerical sign problem. Here, we overcome this obstruction via complex path integration in a geometrically frustrated ladder of interacting bosons at finite density. This enables studies of the many-body ground state properties, otherwise inaccessible with standard quantum Monte Carlo methods. Specifically, we study a chemical potential tuned quantum phase transition, along which we track the emergence of quasi-long-range order and critical softening of the single particle gap. We chart future methodological improvements and applications in generalized geometrically frustrated lattice models.
\end{abstract}
\maketitle

\section{Introduction}
A paradigmatic instance of the numerical sign problem, in the context of condensed matter physics, is geometrically frustrated antiferromagnets. The appearance of non-positive (or even complex) quantum amplitudes renders numerical calculations via Monte Carlo techniques uncontrolled, with statistical errors that scale exponentially with system size and overwhelm the signal. Geometrical frustration enhances quantum fluctuations, promoting exotic and inherently quantum phenomena. Examples thereof include valance bond solids comprising a spatially ordered pattern of singlet dimers~\cite{auerbach1998interacting}, unconventional magnetic textures \cite{tokura2014multiferroics}, and most remarkably, spin liquids that defy ordering down to absolute zero temperature \cite{Savary:2016ksw}. It is, therefore, desirable to devise novel methodologies that overcome the obstruction imposed by the numerical sign problem and provide an accurate numerical solution to geometrically frustrated quantum spin models.

More broadly, a generic solution to the numerical sign problem is likely unfeasible. In fact, in some instances, no-go theorems preclude a complete elimination of the sign problem via local transformations~\cite{Troyer_2005, Hastings_2016, Ringel_2017, Smith:2020, Golan:2020kyp}. Nevertheless, tremendous progress has been made in devising clever reformulations of the path integral representation, providing either a partial or complete elimination of the sign problem in specific models~\cite{Chandrasekharan_1999, Alet_2016, Honecker_2016, Hong_DQMC_free_2019, berg_2012}. Promising recent progress in controlling the numerical sign problem is the complex path integration (CPI) approach~\cite{Alexandru:2015sua, Alexandru:2020wrj}. In this method, the path integral is deformed into the complex plane. An informed choice of the integration manifold can then achieve a significant reduction in the severity of the numerical sign problem. Indeed, numerically accurate investigations of a wide range of many-body problems have been demonstrated, involving both bosonic~\cite{Alexandru:2016san, Alexandru:2016gsd, Alexandru:2017lqr, Bursa:2018ykf, Giordano_2022}, fermionic~\cite{Alexandru:2015xva, Alexandru:2016ejd_GTM, Alexandru:2018ngw, Alexandru:2018ddf}, and spin degrees~\cite{Mooney2021esz} of freedom. 

\begin{figure}[t!]
  \hspace*{-0.4cm}  
    \captionsetup[subfigure]{labelformat=empty}
    \subfloat[\label{subfig:phase_diag}]{}
    \subfloat[\label{subfig:sign_vs_mu}]{}
    \centering
    \includegraphics[scale=0.48]{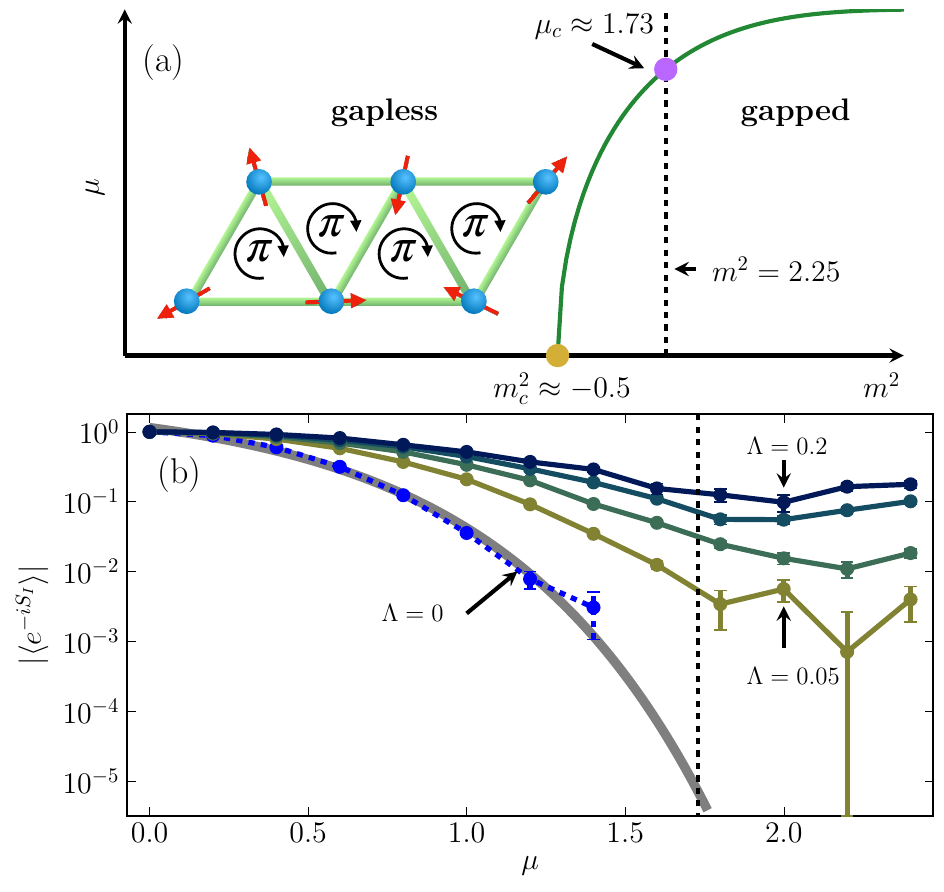}
    \caption{ (a) Illustration of the frustrated ($\pi$-flux) bosonic triangular chain model of \cref{eq:hamiltonian_scalar_bose_gas} and a sketch of the corresponding global phase diagram as a function of the boson mass $m^2$ and chemical potential $\mu$. The yellow dot marks the quantum phase transition at zero boson density ($\mu=0$) between gapped and gapless phases. (b) The average sign, $|\av{e^{-iS_I}}|$, of configuration weights as a function $\mu$ along the chemical potential tuned condensation transition, as marked by the dashed black line and purple dot in (a). Different curves correspond to an increasing range of flow times $\Lambda$, see \cref{eq:holomorphic_flow}. The gray line extrapolates the sign problem along the original integration manifold $\mathbb{R}^{2N}$ to large $\mu$. The vertical dashed line corresponds to the estimated critical chemical potential $\mu_c \approx 1.73$, for $t/|t|=-1$ and $U=1$.}
    \label{fig:fig1}
\end{figure} 

While the CPI approach is a promising avenue for mitigating the numerical sign problem, its overarching applicability in various quantum many-body problems remains an open and timely question. In practice, addressing this question requires a case-by-case study since it is difficult to \textit{a priori} determine the sign structure associated with many-body quantum amplitudes deformed into the complex plane. Specifically, concerning this work, the utility of the CPI to the important problem of geometrically frustrated quantum many-body systems and associated quantum critical behavior is unknown and serves as the central motivation for this study.  

In this work, we address the above inquiry in a geometrically frustrated triangular chain of lattice bosons at a finite chemical potential, a setting for which the sign problem plagues standard quantum Monte Carlo (QMC) methods. Remarkably, we find that integration along complex plane manifolds allows taming the numerical sign problem, leading to controlled numerical calculations. In particular, we track with high precision a chemical potential tuned order-disorder quantum many-body phase transition and analyze the associated finite system size and finite temperature scaling of pertinent physical observables and the numerical sign problem.

\section{Microscopic model and phase diagram} As a concrete lattice model for testing the applicability of the CPI in the context of geometrically frustrated many-body quantum systems, we consider a bosonic lattice model defined on a triangular chain with $L$ sites; see \cref{subfig:phase_diag}. The dynamics is governed by the Hamiltonian,
\begin{equation}
\begin{split}
    \mathcal{H} &= \frac12 \sum_r  \Pi_r^*\Pi_r -t \sum_{\expval{ r,r'}}\psi_{r}^*\psi_{r'} + \text{h.c.} + \sum_r  m^2\qty|\psi_r|^2\\
    &+ \sum_r  U \qty|\psi_r|^4
    + i\mu \sum_r \qty(\psi_r\Pi_r^*-\psi_r^*\Pi_r).
\end{split}
\label{eq:hamiltonian_scalar_bose_gas}
\end{equation}
Here, the complex scalar field operators, $\psi_r$, reside on lattice sites, $r$. We employ linear indexing of sites along an effective one-dimensional chain that alternates between the upper and lower legs of the ladder. The canonical momenta, $\Pi_r$, follows the standard commutation relations $\qty[\psi_r,\Pi_{r'}]=i\delta_{r,r'}$. Complementary relations apply to their complex counterparts, $\psi^*$ and $\Pi^*$. Hamiltonian terms in the first line correspond to the quadratic (``free") part, comprising nearest-neighbor hoppings with amplitude $t$ and a mass term of magnitude $m^2$. The second line includes on-site quartic repulsive interactions of strength $U$ and a chemical potential term. The Hamiltonian admits a global $U(1)$ symmetry corresponding to particle number conservation, as defined by the boson charge operator, $Q=\sum_r \qty(\psi_r\Pi_r^*-\psi_r^*\Pi_r)$.  For nonvanishing $\mu$, particle-hole symmetry, $\psi\to\psi^*$, is broken, which potentially induces a finite particle density, $\expval{Q}\ne 0$.

Making contact with physical systems, we note that our model serves as a low energy effective description of various condensed matter systems with a conserved $U(1)$ charge and broken particle-hole symmetry; see Ref.~\cite{sachdev_2011}. Two concrete examples are lattice bosons detuned from integer filling and easy plane (XY)~magnets in an external magnetic field perpendicular to the magnetization axis. Our primary interest will be in cases where the hopping amplitude is negative, i.e., $t<0$. On nonbipartite lattices, this choice leads to geometric frustration akin to antiferromagnetic interactions in lattice spin models. By contrast, the more standard case, $t>0$, can be interpreted as ferromagnetic interactions and is commonly used in lattice regularizations of continuum quantum field theories \cite{Kogut:1979wt}.

For numerical simulations via QMC techniques, we evaluate the thermal partition function following the standard quantum to classical mapping, $\mathcal{Z}=\int\mathcal{D}\psi \mathcal{D}\psi^* e^{-\mathcal{S}\qty[\psi_{r,\tau},\psi^*_{r,\tau}]}$, that sums over space-time histories of the complex scalar field $\psi_{r,\tau}$, where $\tau$ denotes the imaginary time axis. The corresponding action reads, 
\begin{equation}
\begin{split}
    S &= 
    -\varepsilon t \sum_{\expval{r,r'},\tau} (\psi_{r,\tau}\psi_{r',\tau}^* + \text{h.c.}) \\
    &- \frac{1}{2\varepsilon} \sum_{r,\tau}
    \qty(
    \psi_{r,\tau}^*\psi_{r,\tau+\varepsilon}\qty(1-\varepsilon\mu) 
   +\psi_{r,\tau}\psi_{r,\tau+\varepsilon}^*\qty(1+\varepsilon\mu)
    )\\
    &+\sum_{r,\tau} \qty[
    \qty(\varepsilon m^2 -\frac{\varepsilon\mu^2}{2} + \frac{1}{\varepsilon}) \qty|\psi_{r,\tau}|^2 
    + \varepsilon U \qty|\psi_{r,\tau}|^4
    ].
\end{split}
\label{eq:action_scalar_bose_gas}
\end{equation}
In the above equation, the Trotter step equals $\epsilon=\beta/L_\tau$, with $\beta$ denoting the inverse temperature, and $L_\tau$ is an integer that defines the length of the discrete imaginary time axis.
Notably, the action involves complex weights for a finite $\mu$, rendering direct Monte Carlo sampling uncontrolled due to the notorious numerical sign problem. 

Interestingly, for $t>0$, the path integral can be reformulated in terms of the so-called dual variables \cite{Prokofev_1998, Endres_2007}, for which configuration weights are real and strictly nonnegative. Physically, this representation tracks boson world line configurations, such that the case $\mu>(<)0$ favors particles (holes) worldlines propagating along the positive (negative) imaginary time axis. By contrast, for negative hopping amplitudes, $t<0$, on nonbipartite lattices, boson world lines acquire a $\pi$ phase associated with trajectories encircling an odd number of bonds; see \cref{subfig:phase_diag}. This effect reintroduces the numerical sign problem and addressing it using CPI is the main focus of this work.

We now briefly chart the zero-temperature phase diagram of our model [\cref{eq:hamiltonian_scalar_bose_gas}] in its limiting cases. For a vanishing chemical potential, $\mu=0$, a quantum phase transition separates a gapped phase for large and positive mass, $m^2$, from a gapless phase with quasi-long-range order (QLRO) in the opposite limit of large and negative $m^2$. The transition belongs to the Berezinskii–Kosterlitz–Thouless (BKT) universality class and occurs at a critical coupling $m^2_c$. We note that, due to the $\pi$-flux pattern, for negative hopping amplitudes, $t<0$, QLRO correlations develop an incommensurate spiral pattern at finite Bragg wave vector $\tilde{q}=\cos^{-1}(-1/4)$; see \cref{subfig:phase_diag} and \cref{app:mean_field_sol_and_mu_0}.

For a given hopping amplitude, starting from the disordered phase, $m^2>m^2_c$, the single particle gap can also be closed by increasing the chemical potential $\mu$. Physically, this induces a BEC-like transition, where the chemical potential provides the necessary energy for particles to overcome the gap and condense~\cite{Sachdev_1994}. However, unlike the $\mu=0$ transition, here, the transition is nonrelativistic with a dynamical critical exponent $z=2$. In the context of matter fields at finite density, such transitions are commonly termed the ``Silver Blaze" effect \cite{Cohen:2003kd}.

\section{Numerical methods and observables} We now briefly review the construction of our CPI scheme using the generalized thimble method (GTM) \cite{Alexandru:2016san, Alexandru:2015sua, Alexandru:2016lsn_wrongian, Alexandru:2017oyw_pt, Alexandru:2020wrj}. Within this approach, the complex plane integration manifold is determined through the holomorphic flow equation, 
\begin{equation} \label{eq:holomorphic_flow}
    \dv{\psi_{r,\tau}}{\lambda} = \overline{\pdv{S}{\psi_{r,\tau}}},
\end{equation}
where $\lambda$ parametrizes the flow time. We set $\mathbb{R}^{2N}$ as the initial condition, of the above differential equation, at flow time $\lambda=0$, which is associated with the real and imaginary parts of the complex fields $\psi_{r,\tau}$ residing on $N=L\times L_\tau$ space-time points. The equation is then integrated up to a flow time $\lambda=\Lambda$, which induces a mapping between $\mathbb{R}^{2N}$ (the original integration manifold) and $\mathcal{M}_\Lambda$, which is embedded in $\mathbb C^{2N}$.

This construction is motivated by the limiting manifolds at $\Lambda\to\infty$, known as the Lefschetz thimbles \cite{Alexandru:2020wrj}. Importantly, along each Lefschetz thimble, the imaginary part of the action is constant, which, at least formally, eliminates the numerical sign problem. However, the Lefschetz thimble structure may fracture into multiple disconnected thimbles that assign potentially distinct phases to their quantum amplitudes in the complex plane. Interference between different thimbles can then reintroduce the numerical sign problem. The flow time $\Lambda$ presents a trade-off between reducing the numerical sign problems at long flow times versus ergodicity issues in the Monte Carlo dynamics arising from the trapping of MC configurations in the vicinity of a specific thimble. To address this problem, we employ the parallel tempering technique that exchanges between configurations at varying flow times $\Lambda$. This allows for smooth interpolation between distinct thimbles \cite{Fukuma:2017fjq, Alexandru:2017oyw_pt}.  Residual phases of configuration weights are considered through the standard reweighting approach.

\begin{figure}
  \hspace*{-0.5cm}  
    \captionsetup[subfigure]{labelformat=empty}
    \subfloat[\label{subfig:sign_vs_beta_L}]{}
    \subfloat[\label{subfig:Q_vs_mu_beta}]{}
    \centering
    \includegraphics[width=0.49\textwidth]{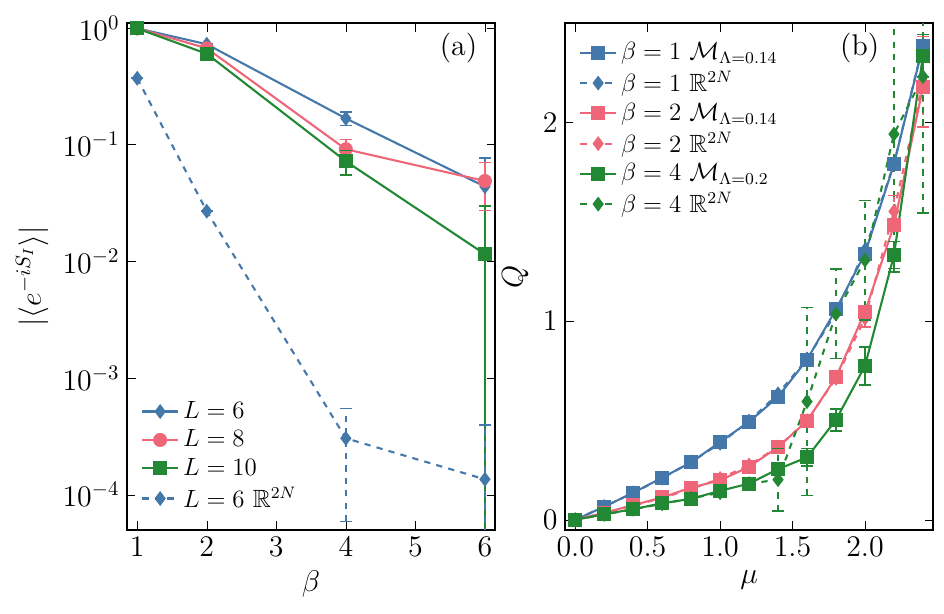}
    \caption{(a) Average sign, $\qty|\av{e^{-iS_I}}|$, evaluated at $\mu=1.6$, as a function of the inverse temperature $\beta$. Solid curves correspond to flow time $\Lambda=0.2$. The dashed line represents a vanishing flow time, $\Lambda=0$. Different curves are associated with different system sizes. (b) The average particle number $Q$ as a function of $\mu$ for $L=6$. Different curves correspond to different inverse temperatures. Solid (dashed) lines depict finite (vanishing) flow times.}
    \label{fig:fig2}
\end{figure}

We now turn to defining physical observables, characterizing the various phases and phase transitions appearing in our problem. We first consider the particle number density $Q=\frac1{L\beta}\pdv{\ln{\mathcal{Z}}}{\mu}$, explicitly given by 
\begin{equation}\label{eq:charge}
    Q=\frac1{L\beta}\expval{\sum_{r,\tau}\frac{1}{2} \qty(
    \psi_{r,\tau}^*\psi_{r,\tau+\varepsilon}
   -\psi_{r,\tau+\varepsilon}^*\psi_{r,\tau})
+\varepsilon\mu  \qty|\psi_{r,\tau}|^2 },
\end{equation}
which tracks the breaking of particle-hole symmetry.
To probe the evolution of space-time correlations, we compute the single particle Green's function,
$G(q,\omega_m)=\frac{1}{L\beta}\expval{\qty|\int_0^\beta d\tau\sum_{r}\psi_{r,\tau}e^{i\qty(qr+\omega_m\tau)}|^2}$.
Here, $\omega_m=2\pi m/\beta$, are the standard bosonic Matsubara frequencies for integer $m$ and integration along the imaginary time axis is discretized, as defined above. The corresponding imaginary time correlations, $G(q,\tau)$, are obtained by Fourier relations.

With the above definitions, in the condensed phase, we expect to find QLRO, which we detect by examining the equal time Green's function evaluated at the Bragg vector $g(\bar{q})=G(q=\bar{q},\tau=0)$, with $\bar{q}$ taken as the closest approximate to $\tilde{q}$ on our finite-size lattice. 

The low energy dynamics is studied through the expected long imaginary time exponential decay,
\begin{equation}
    G(\tilde{q},\tau>0) \sim e^{-\Delta_{\tilde{q}}\tau}
\end{equation}
We estimate the single particle gap for particles,  $\Delta_{\tilde{q}}$ by fitting to the above form. The gap for antiparticles (holes) can be extracted similarly from the decay at negative times $\tau<0$. 

\begin{figure*}[t!]
\hspace*{-0.6cm}  
    \centering
    \captionsetup[subfigure]{labelformat=empty}
    \subfloat[\label{subfig:Q_vs_mu}]{}
    \subfloat[\label{subfig:gmax_vs_mu}]{}
    \subfloat[\label{subfig:Delta_vs_mu}]{}
    \includegraphics[width=1.00\textwidth]{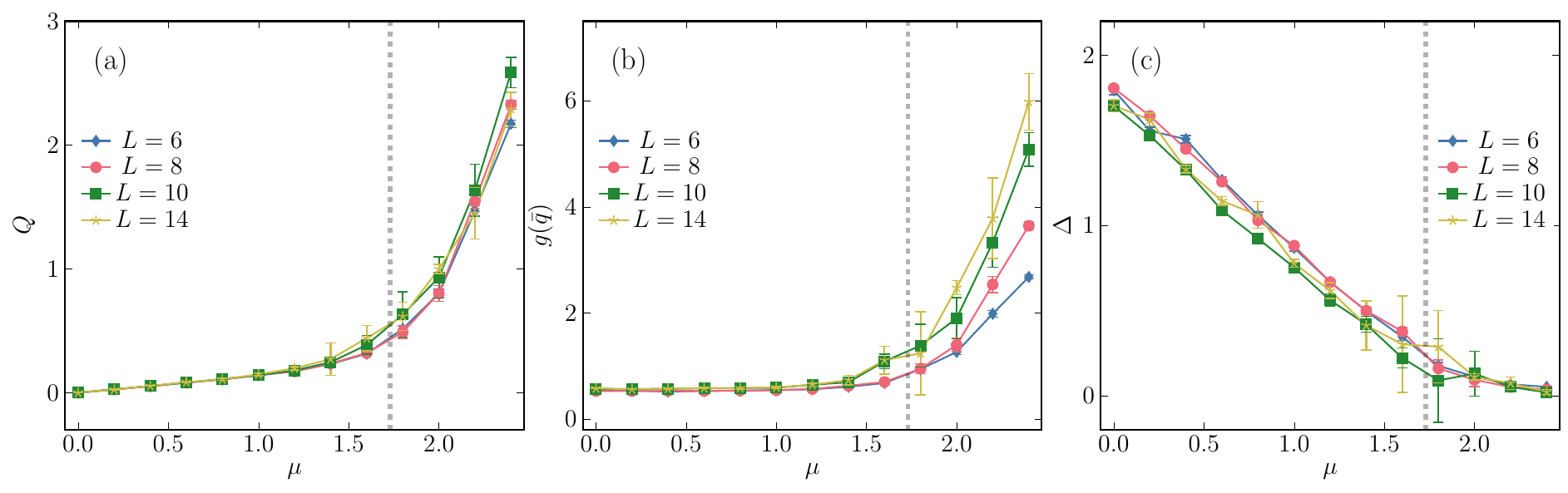}
    \caption{Chemical potential tuned quantum phase transition.~(a) Boson particle number density, $Q$, (b) equal time Green's function at the Bragg vector $g(\bar{q})$, 
    and (c) the single particle gap $\Delta$. 
    Different curves correspond to increasing values of system size $L$. Simulations were carried out at inverse temperature $\beta=4$. The flow time for $\mu<1.4$ equals $\Lambda=0.15$ and for $1.4 \le \mu \le 2.4$, $\Lambda=0.2$. The vertical line at $\mu=1.73$ marks the approximate position of the phase transition obtained from vanishing chemical potential simulations.}
    \label{fig:fig3}
\end{figure*}

\section{Numerical results} For concreteness, we measure all energy scales in units of $\sqrt{\abs{t}}$ and fix ${U=1},{m^2=2.25}$, and ${t/|t|=-1}$. For this parameter choice and a vanishing chemical potential $\mu=0$, we are located in the disordered (gapped) phase, $m^2>m^2_c\approx-0.5$, see \cref{subfig:phase_diag} and \cref{app:mean_field_sol_and_mu_0}. For finite size and finite temperature analysis, we consider system sizes $L=6,8,10,14$ and track the convergence to the ground state result by progressively increasing the inverse temperature, taking the values $\beta=1,2,4,6$. We observed that within our parameter regime, $\beta=4$ serves as a proxy for the ground state behavior. Throughout, the Trotter step equals $\epsilon=0.17$. 

We begin our analysis by probing the evolution of the numerical sign problem as a function of the chemical potential, $\mu$, for an increasing range of the flow times $\Lambda$, as shown in \cref{subfig:sign_vs_mu}. As expected, a naive integration over $\mathbb{R}^{2N}$ displays a hard sign problem upon approach to the critical chemical potential $\mu_c\approx 1.73$, as evident by the rapid drop towards zero of the average sign, $|\av{e^{-iS_I}}|$. Remarkably, examining finite flow times, $\Lambda$, progressively reduces the sign problem even for the challenging parameter regime $\mu>\mu_c$. 

To further substantiate the above result, in \cref{subfig:sign_vs_beta_L}, we examine the average sign for flow time $\Lambda=0.2$, chemical potential $\mu=1.6$, and an increasing range of system sizes and inverse temperatures. Crucially, at low temperatures and for all $L$ values, the average sign along $\mathcal{M}_\Lambda$ rises in almost three orders of magnitude compared with the one computed on $\mathbb{R}^{2N}$ for $L=6$. This key finding is one of our main results, which enables an accurate numerical study of our geometrically frustrated model via Monte Carlo sampling, as we demonstrate below.

Turning to physical observables, we test the advantage of finite flow times ($\Lambda>0$) simulation against ``brute force" integration on $\mathbb{R}^{2N}$. To that end, we measure the average particle number $Q$ as a function of the chemical potential $\mu$ for $\beta=1,2,4$. In all cases, due to the severe sign problem, we considered a factor of 128 times more Monte Carlo samples for $\mathbb{R}^{2N}$ integration than the finite flow time simulations. The results of this analysis are shown in \cref{subfig:Q_vs_mu_beta}. Indeed, at low temperatures, $\beta=4$ converged results are only obtained using finite flow times, demonstrating the utility of the GTM sampling. Even more impressively, this advantage is obtained despite the significantly reduced MC samples. We note that, for larger system sizes, direct comparison is completely infeasible in realistic run times due to the vanishingly small average sign in $\mathbb{R}^{2N}$ simulations.

After establishing control over the numerical sign problem within the parameter range of interest, we investigate the physical properties of our many-body problem in the vicinity of the chemical potential tuned quantum phase transition described above. In the following, we fix $\beta=4$, and explore the finite size scaling properties with $L=6,8,10$ and $14$. First, we compute the evolution of the particle number, $Q$, as a function $\mu$, in \cref{subfig:Q_vs_mu}. As expected, we find that, at small $\mu$, the particle number vanishes and starts to increase only above a critical coupling, $\mu_c\approx1.73$, consistent with the estimate of the single particle gap computed at $\mu=0$; see \cref{app:mean_field_sol_and_mu_0}. 

Next, we track the order parameter $g(\bar{q})$ as a function of $\mu$, as depicted in \cref{subfig:gmax_vs_mu}. We observe a rise of $g(\bar{q})$ for $\mu>\mu_c$, signaling the appearance of QLRO. 
Lastly, in \cref{subfig:Delta_vs_mu}, we address dynamical properties by computing the gap for particle excitations $\Delta$. The numerical results agree with the predicted gap closing transition at $\mu_c$, which nucleates the condensed phase.

\section{Discussion and summary} In this work, we have demonstrated the effectiveness of the CPI approach in controlling the numerical sign problem appearing in a geometrically frustrated quantum many-body system. In particular, we identify complex plane manifolds, $\mathcal{M}_\Lambda$, over which the severity of the sign problem is progressively reduced as a function of the flow time $\Lambda$. This methodological headway enabled access to an accurate numerical study of collective effects in the vicinity of a quantum critical point.

Our work primarily concerns ladder geometries. In such cases, tensor network approaches, such as the density matrix renormalization group and tensor renormalization \cite{White_dmrg_1992, Vidal_trg_2015} that target low-entangled states can provide an efficient and accurate numerical solution. However, beyond one dimension and even worse in the proximity of quantum criticality, the unbounded growth of the entanglement entropy hinders the applicability of these methods. Our results indicate that, at least for the restricted ladder geometry, CPI also allows a sizable reduction in the severity of the numerical sign, enabling controlled numerical calculation. This key observation serves as an encouraging first step and motivates extensions to a higher dimension. As a proof of principle, in \cref{app:two_dimension}, we showcase the applicability of the CPI approach for a small ($3\times3$) two-dimensional frustrated triangular lattice. We, indeed, overcome the inherent numerical sign problem and observe a chemical potential-driven transition towards a 120$^\circ$ ordered phase.

Looking to the future, despite a great deal of progress, standard application of the GTM approach remains computationally demanding, mainly due to the repeated numerical solution of the holomorphic flow equation \cref{eq:holomorphic_flow}. In that regard, it would be beneficial to explore optimization techniques over families of analytically defined complex plane manifolds~\cite{Alexandru:2018fqp} or machine learning based approaches for constructing integration manifolds~\cite{Alexandru:2017czx}.

From the physics front, our results open the door to studies of more involved geometrically frustrated lattice models. Natural extensions include approaching the hard-core limit $U \gg t$, and the more audacious goal of addressing the two-dimensional triangular lattice \cite{Zaletel_2014} in the thermodynamic limit. We leave these exciting research directions to future studies.

\begin{acknowledgments}
We thank A. Auerbach, E. Berg, and Z. Ringel for helpful discussions and C. Rackauckas for support in employing the {\it 
DifferentialEquations.jl} package. 
S.G. acknowledges support from the Israel Science Foundation (ISF) Grant No. 586/22 and the US–Israel Binational Science Foundation (BSF) Grant No. 2020264. E.C. acknowledges support from an anonymous donor from the United Kingdom. A.A. is supported in part by U.S. DOE Grant No. DE-FG02-95ER40907. This research used the Intel Labs Academic Compute Environment.
\end{acknowledgments}

\appendix

\section{Details of the Monte Carlo simulation}
\label{Ap:MC}
In our numerical simulations, we employed the generalized thimble method (GTM) \cite{Alexandru:2016ejd_GTM}. The key idea of this approach is to construct a complex plane integration manifold via a mapping of $\mathbb{R}^{2N}$ as defined by the holomorphic flow equation up to time $\Lambda$. The motivation for this particular choice is derived from the long flow time limit $\Lambda\to\infty$, for which the integration manifold coincides with the set of Lefschetz thimbles. Since, by definition, the complex part of the action is stationary on each Lefschetz thimble, the effect of the numerical sign problem is anticipated to be diminished. We note that different thimbles may carry a distinct phase and hence lead to an interference effect. Therefore, the GTM approach is expected to be advantageous when a small number of thimbles dominate the contribution to path integral. 

From the technical perspective, to speed up the computation of the Jacobian associated with the mapping between $\mathbb{R}^{2N}$ and $\mathcal{M}_\Lambda$, we use the Wrongian~\cite{Alexandru:2016lsn_wrongian} approximation for the Jacobian and correct via reweighting. For the numerical integration of the holomorphic flow \cref{eq:holomorphic_flow} in the main text, we employed the \textit{DifferentialEquations.jl} package \cite{rackauckas2017differentialequations} using the Tsit5 algorithm \cite{tsitouras2011runge}. We allow for a local integration error tolerance of $10^{-3}$, which we empirically found to be sufficient for obtaining converged results.

In some instances, the long flow time integration manifold is fractured into multiple thimbles separated by large energy barriers. Consequently, the Monte Carlo sampling may turn nonergodic due to the low acceptance rate of trajectories connecting distinct thimbles. To address this issue, we consider a parallel tempering (PT) scheme \cite{Fukuma:2017fjq, Alexandru:2017oyw_pt} for a linearly spaced range of flow times $\Lambda_i=\{0,\Delta \Lambda,2\Delta \Lambda,\ldots, \Lambda_{max}\}$. Typically, we considered 15--25 walkers and averaged our results on the last 3--5 walkers. This approach enables a smooth interpolation between disjoined thimbles at long flow times via thermalization at short flow times. Importantly, it still retains the essential aspect of suppressing the numerical sign problem when sampling from the long flow time manifolds. To monitor the efficiency of the PT scheme, we used the methods described in \cite{Katzgraber_2006} and found a sufficiently high rate swap move between PT replicas. To initialize the field configurations, we used simulated annealing by progressively increasing the flow time $\Lambda$ in the thermalization phase of the Monte Carlo sampling.

\begin{figure}[b]
    \centering
    \includegraphics[width=\linewidth]{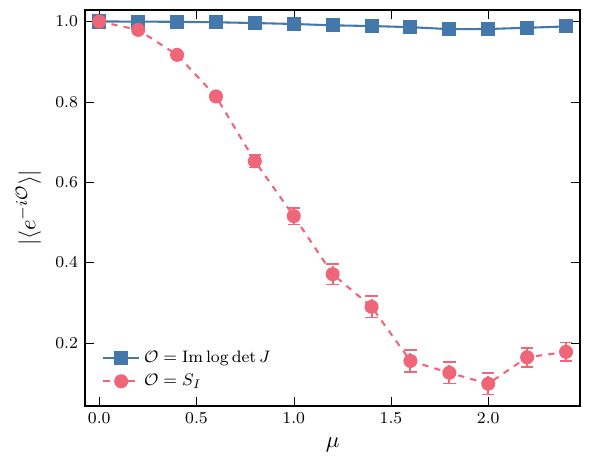}
    \vspace{10pt}
    \caption{Comparison of the average phase emanating from the imaginary part of the action and the imaginary part of the coordinate transformation Jacobian determinant ($\log\det J$). Simulations are carried out at $L=6,\beta=4, \Lambda=0.2$, and the rest of the parameters are the same as the main text.}
    \label{fig:jacobian_phase}
\end{figure}
\begin{figure}[t]
    \captionsetup[subfigure]{labelformat=empty}
    \centering
    \includegraphics[width=\linewidth]{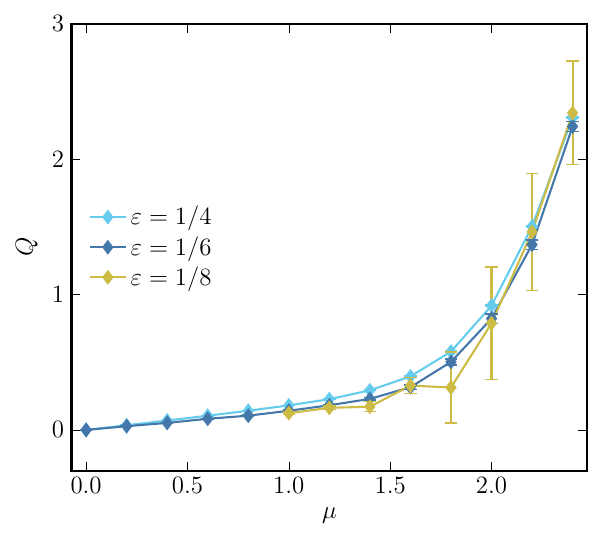}
    \vspace{10pt}
    \caption{Particle number density $Q$ for $\beta=4$ and $L=6$, as a function of the chemical potential $\mu$. The rest of the microscopic parameters are the same as the main text. Different curves correspond to different values of $\varepsilon$. }
    \label{fig:SM_epsilon_convergence}
\end{figure}
\begin{figure}
    \captionsetup[subfigure]{labelformat=empty}
    \subfloat[\label{subfig:particle_CPI_vs_worm}]{}
    \subfloat[\label{subfig:green_CPI_vs_worm}]{}
    \centering
    \includegraphics[width=\linewidth]{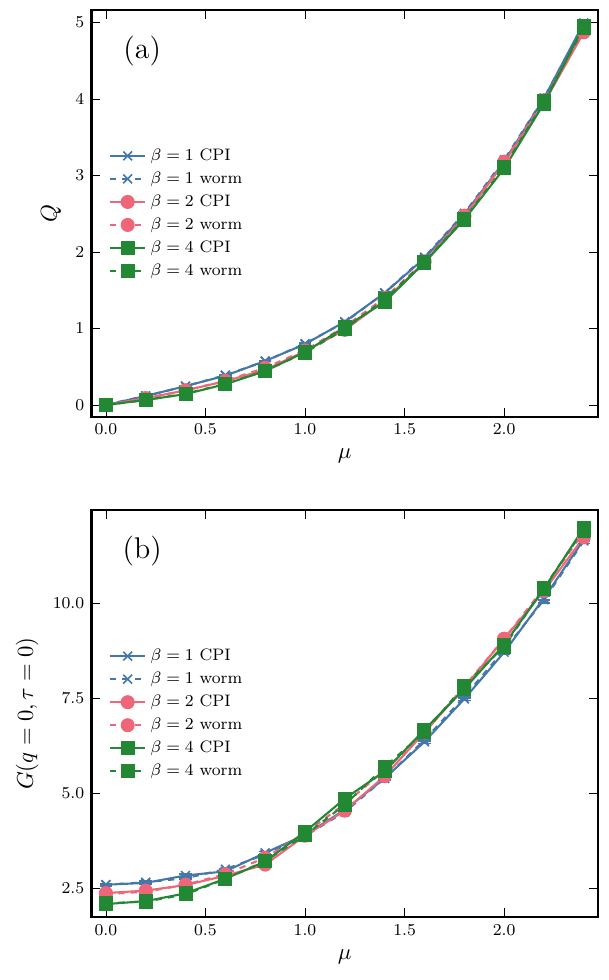}
    \caption{Comparison of the GTM algorithm with the worm algorithm for ferromagnetic interaction. We present results for $L=6,m^2=2.25, U=1, t=+1,\varepsilon=1/6$ . For  GTM we take a flow time $\Lambda=0.14$. In (a) we depict the average particle number, $Q$, and in (b) the order parameter, $G(q=0,\tau=0)$, both as a function of the chemical potential $\mu$. }
    \label{fig:benchmark_worm}
\end{figure}

\begin{figure}
    \captionsetup[subfigure]{labelformat=empty}
    \subfloat[\label{subfig:SM_green_q_RN_vs_CPI}]{}
    \subfloat[\label{subfig:SM_green_tau_RN_vs_CPI}]{}
    \centering
    \includegraphics[width=\linewidth]{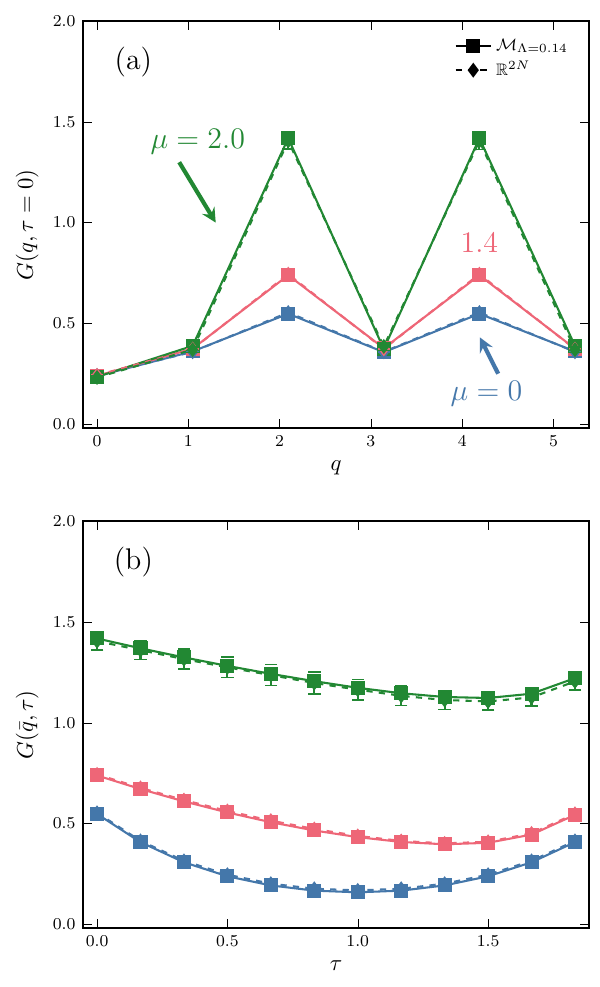}
    \caption{Comparison between $\mathbb{R}^{2N}$ (dashed line) and finite flow time $\Lambda = 0.14$ (solid line) integration for ${L = 6}, {\beta = 2}, {\varepsilon = 1/6}, {U=1}$ and frustrated (antiferromagnetic) hopping $t=-1$. (a)~Equal time Green's function $g(q)$ and~(b)~$G(\bar q, \tau)$. Different curves correspond to a different chemical potential $\mu=0,1.4,2.0$. $\mathbb{R}^{2N}$ simulations are carried out with 128 times more measurements than the simulation at $\mathcal{M}_\Lambda$.}
    \label{fig:benchmark_brute_force}
\end{figure}

We note that the complex amplitudes appearing in our path integral originate both from the standard imaginary part of the action $\Im S$ and the Jacobian determinant associated with the coordinates' transformation of the real plane to the complex plane manifold. Typically, the latter has a minor contribution to the numerical sign problem. Indeed, we find that, also in our case, $\Im\log\det(J)$ is vanishing small. This is exemplified in \cref{fig:jacobian_phase}, where we show that the contribution to the sign problem corresponding to the action $\Im S$ overwhelms the one originating from  $\Im\log\det(J)$.

Lastly, to quantify the effect of Trotter errors, we carry out QMC simulations at $m^2=2.25, U=1, \beta=4$ and $L=6$, using a sequence of decreasing values of $\varepsilon$, see \cref{fig:SM_epsilon_convergence}. Indeed, we confirm that  $\varepsilon\le 1/6$, as used in the main text, is sufficiently small to produce converged results within the statistical errors.

\section{Benchmarking the GTM integration}

To verify the correctness of our numerical implementation of the GTM, we carried out extensive benchmarking, comparing our numerical results with brute force calculations and, when possible, complementary numerically exact methods. 

First, we consider the ferromagnetic case, $t>0$. Here, as mentioned in the main text, by formulating the path integral in terms of dual variables, one can completely eliminate the sign problem. The resulting world lines configuration space is then efficiently sampled via the Worm algorithm~\cite{Prokofev_1998}. We note that the direct representation ($S\qty[\psi,\psi^*]$) still suffers from a severe sign problem. In~\cref{fig:benchmark_worm}, we present the result of this analysis. We find excellent agreement both for the particle number density, $Q$, and the two-point Green's function $G(q=0,\tau=0)$. We note that this nontrivial check also nontrivially corroborates the correctness of our implementation for the frustrated case since, operationally, it only involves changing the sign of the hopping amplitude and does not affect the GTM procedure.

Next, we benchmark the frustrated case, $t<0$. Here, due to the severe sign problem, sampling at vanishing flow times ($\Lambda=0$) requires substantial statistics and hence enables addressing only small system sizes at relatively high temperatures. Concretely, we consider $L=6,\beta=2$. Reassuringly, in \cref{fig:benchmark_brute_force}, we indeed find excellent agreement between the brute force approach and GTM integration, both for $G(q,\tau=0)$ in \cref{subfig:SM_green_q_RN_vs_CPI} and $G(\bar{q},\tau)$ in \cref{subfig:SM_green_tau_RN_vs_CPI}. This result further supports our numerical scheme. We note that despite using 128 times more measurements for $\Lambda=0$ compared to $\Lambda=0.14$, we still obtained results with smaller error bars in the latter case.

\section{Mean field solution and estimation of the gap at \texorpdfstring{$\mu=0$}{mu=0}}
\label{app:mean_field_sol_and_mu_0}

\begin{figure*}
    \captionsetup[subfigure]{labelformat=empty}
    \subfloat[\label{subfig:SM_green_mu0}]{}
    \subfloat[\label{subfig:SM_q_tilde_mu0}]{}
    \subfloat[\label{subfig:SM_Delta_mu0}]{}
    \centering
    \includegraphics[width=\linewidth]{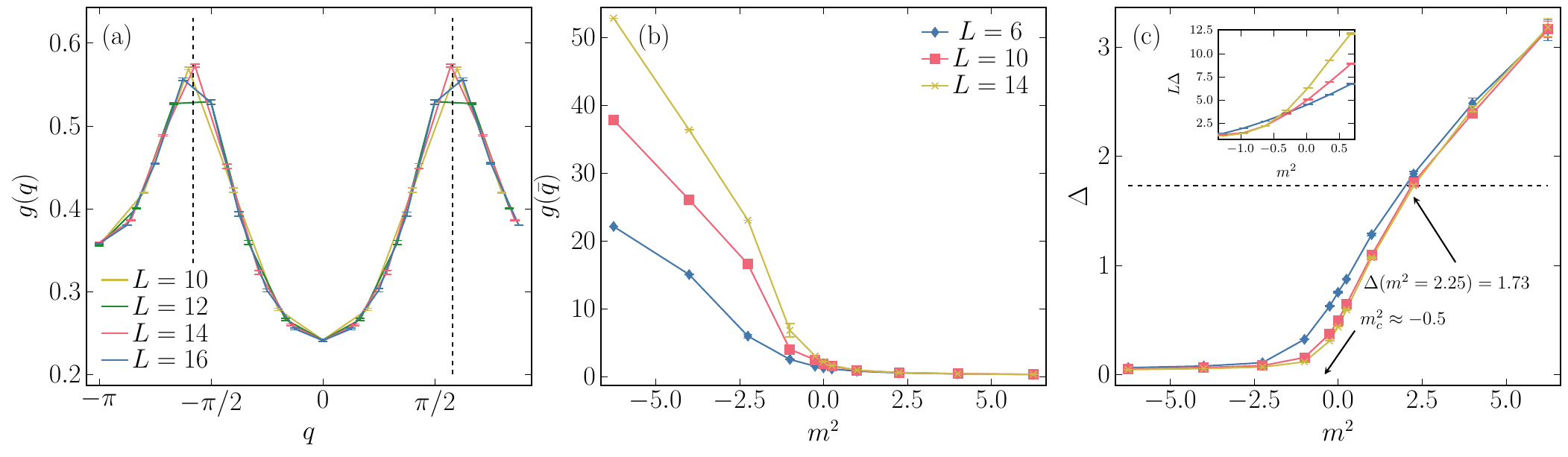}
    \caption{Quantum phase transition at $\mu=0$. We take an inverse temperature $\beta=12$ as a proxy for ground state properties. (a) Equal time Green's function $g(q)$ for different system sizes at $m^2=2.25$. The dashed lines mark the position of $\tilde{q}$. Finite-size scaling across the transition: (b) equal time Green's function evaluated at $\bar q$ as a function of $m^2$; (c) the single particle gap $\Delta$. The horizontal dashed line marks the value of the single particle at $m^2=2.25$, used in the main text. The inset in (c) represents a curve crossing analysis with the dimensionless amplitude $L\Delta$ (the dynamical exponent $z=1$). We identify the phase transition at the crossing point $m^2_c\approx-0.5$.}
    \label{fig:mu0_mean_field}
\end{figure*}

It is illuminating to carry out a simple mean field analysis of our model in the particle-hole symmetric limit ($\mu=0$) in order to predict the ordering pattern in the condensed phase. To that end, we analyze an effective quadratic model corresponding to the disordered phase. In this regime, non-linear terms, for finite $U$, will only renormalize the hopping and mass terms. We then search for the leading susceptibility divergence as we approach the ordered phase. More concretely, we analyze the quadratic action
\begin{equation} \label{eq:action_kernel}
    S = \sum_{q,\omega_m} 
    \qty(\omega_m^2 + m^2-t \qty(\cos(q)+\cos(2q)))
    \psi_{q,\omega}^*\psi_{q,\omega}.
\end{equation}
From the above equation, we isolate the $q$-dependent mass $M^2(q)=m^2-t \qty(\cos(q)+\cos(2q))$. The resulting Green's function displays a peak at momentum $\tilde{q}=\cos^{-1}\qty(-1/4)\approx 0.58\pi$, which corresponds to planar magnetic order with a rotation angle $\theta\approx 104^\circ$ between adjacent spins.

We corroborate this prediction by exact numerical simulations, which for $\mu=0$ are free of the numerical sign in the direct ($S[\psi,\psi^*]$) representation. Concretely, we scan the phase diagram as a function of $m^2$; all other parameters appear as in the main text. In \cref{subfig:SM_green_mu0}, we depict the equal-time Green's function $g(q)$, in the gapped phase, $m^2=2.25$. Indeed, we find a maximum in the vicinity of $\tilde{q}$, indicating the incipient quantum critical point. Next, we drive the transition by further lowering $m^2$. As predicted, we observe a divergence of $g(\tilde{q})$, in \cref{subfig:SM_q_tilde_mu0}, and vanishing of the single particle gap, $\Delta$, in \cref{subfig:SM_Delta_mu0}, at a critical value $m_c^2\approx-0.5$. From this analysis, we also estimate the gap at $m^2=2.25$ to be $\Delta=1.73$. We use this value in the main text to estimate the critical chemical potential $\mu_c$.

\section{Two-dimensional simulation of a triangular lattice}
\label{app:two_dimension}

\begin{figure*}
    \captionsetup[subfigure]{labelformat=empty}
    \subfloat[\label{subfig:SM_2d_sign}]{}
    \subfloat[\label{subfig:SM_2d_Q}]{}
    \subfloat[\label{subfig:SM_2d_q_tilde}]{}
    \centering
    \includegraphics[width=\linewidth]{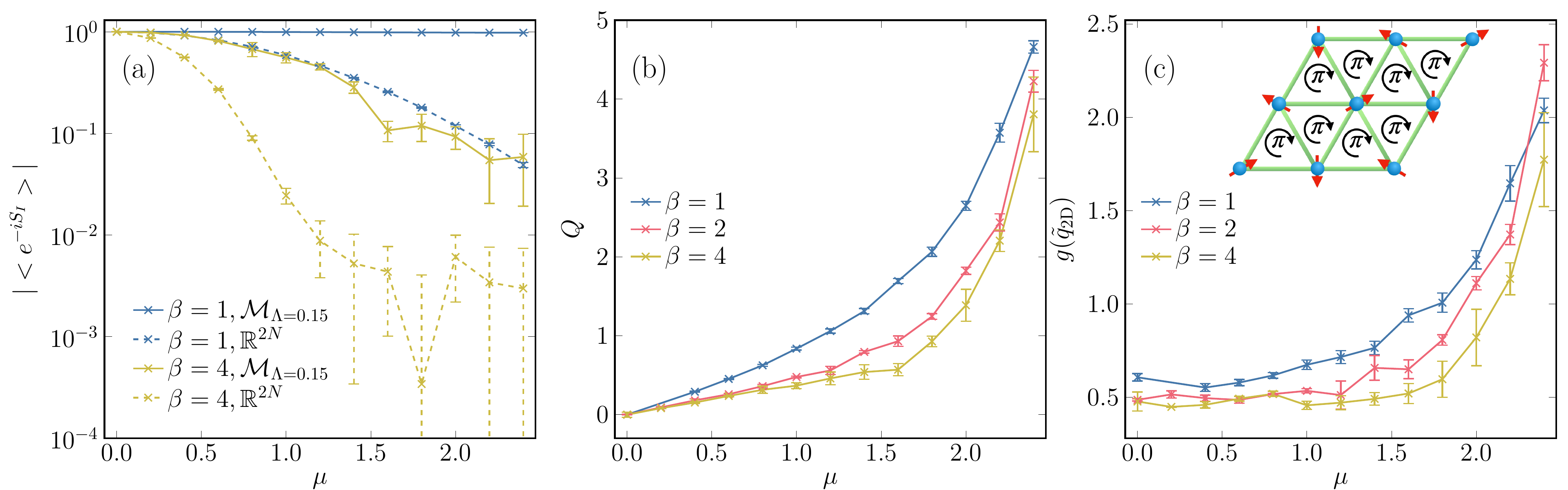}
    \caption{Simulations of a $3\times3$ two-dimensional frustrated triangular lattice. (a) The average sign for an increasing range of $\mu$ and inverse temperatures $\beta=1$ and 4. Solid (dashed) lines correspond to integration along $\mathcal{M}_{\Lambda=0.15}(\mathbb{R}^{2N})$. The particle number density (b) and the structure factor (c) at the Bragg vector $\tilde{q}_\text{2D}$ for $\beta=1,2,$ and 4 as a function of $\mu$. In all cases, we set $\varepsilon= 1/6, m^2=4,t/\abs{t}=-1, U=1$. In the inset, we illustrate the 120$^\circ$ pattern that emerged in the ordered phase.}
    \label{fig:SM_2d}
\end{figure*}

In this section, we investigate the applicability of the CPI method in small two-dimensional systems. More concretely, we consider an extension of the model, appearing in \cref{eq:holomorphic_flow} of the main text to the full two dimensions triangular lattice. For concreteness, we fix antiferromagnetic interactions $t/\abs{t}=-1$ and set $m^2=4, U=1$, $\varepsilon=1/6$, such that for vanishing chemical potential we are in the gapped phase. We study a $3\times3$ triangular lattice with periodic boundary conditions and we probe the convergence into the ground state with inverse temperature $\beta=1,2,$ and $4$. By tuning the chemical potential, we induce an ordering transition that is expected to develop at the Bragg vectors $\tilde{q}_{\text{2D}}=(2\pi/3,2\pi/\sqrt3),(-2\pi/3,2\pi/\sqrt3),(4\pi/3,0)$ corresponding to 120$^\circ$ order \cref{subfig:SM_2d_q_tilde}. For the CPI simulation, we take the flow time $\Lambda=0.15$.

Indeed, we find a significant reduction in the severity of the sign problem, up to one order of magnitude. This result is shown in \cref{subfig:SM_2d_sign} by comparing simulations with finite and vanishing flow times. Encouraged by these results, we probe the ordering transition by computing the average particle number density, $Q$, and structure factor $g(\tilde{q}_{\textrm{2D}})$, at the Bragg vector. The simulation results are consistent with an ordering transition at $\mu\approx2.0$ (\cref{subfig:SM_2d_Q} and \cref{subfig:SM_2d_q_tilde}).

\bibliography{Frustrated_Thimbles} 

\end{document}